\documentclass[11pt]{article}
\usepackage{graphicx} 

\usepackage{graphicx} 
\usepackage{amsmath,amssymb} 

\usepackage[numbers,sort&compress]{natbib}

\usepackage{geometry}
\usepackage{hyperref}
\usepackage{multirow}
\usepackage{enumitem}

\usepackage{soul}
\usepackage{color}
\title{Synergies between Roman Galactic Plane Survey\\ and other major surveys}
\date{}
\begin{document}

\maketitle

\noindent{\bf Scientific Categories: } stellar physics and stellar types; stellar populations and the interstellar medium.\\

\noindent{\bf Additional scientific keywords: } Astrometry; Black holes; Star formation; Exoplanets;

\noindent{\bf Submitting Author: }
Katarzyna Kruszy\'{n}ska, Las Cumbres Observatory, kkruszynska@lco.global\\

\noindent{\bf List of contributing authors: }\\
Rachel A. Street, Las Cumbres Observatory,\\ 
Steven Gough-Kelly, Jeremiah Horrocks Institute, University of Central Lancashire, \\ 
Rosaria (Sara) Bonito - INAF/Osservatorio Astronomico di Palermo (Italy), \\ 
Loredana Prisinzano - INAF/Osservatorio Astronomico di Palermo (Italy), \\ 
Oem Trivedi - ICSC Ahmedabad University,\\ 
Poshak Gandhi, University of Southampton (UK), \\ 
Markus Hundertmark, Astronomisches Rechen-Institut, Zentrum f{\"u}r Astronomie der Universit{\"a}t Heidelberg (Germany), \\ 
Yiannis Tsapras, Astronomisches Rechen-Institut, Zentrum f{\"u}r Astronomie der Universit{\"a}t Heidelberg (Germany),\\
Marcella Di Criscienzo, INAF-Osservatorio Astronomico di Roma (Italy), \\ 
Ilaria Musella, INAF-Osservatorio Astronomico di Capodimonte (Italy),\\
Massimo Dall'Ora, INAF-OACN (Italy), \\
Etienne Bachelet, IPAC, Mail Code 100-22, Caltech, 1200 E. California Blvd., Pasadena, CA 91125, USA,\\ 
Natasha S. Abrams, University of California, Berkeley \\
Somayeh Khakpash, Rutgers University, NJ, \\ 
Markus Rabus, Departamento de Matem{\'a}tica y F{\'i}sica Aplicadas, Facultad de Ingenier{\'i}a, Universidad Cat{\'o}lica de la Sant{\'i}sima Concepci{\'o}n, Alonso de Rivera 2850, Concepci{\'o}n, Chile,\\ 
Paula Szkody, University of Washington, USA,\\ 
Carrie Holt, Las Cumbres Observatory,\\ 

\section{Abstract}
Nancy Grace Roman Space Telescope will revolutionize our understanding of the Galactic Bulge with its Galactic Bulge Time Domain survey. At the same time, Rubin Observatories's Legacy Survey of Space and Time (LSST) will monitor billions of stars in the Milky Way. 
The proposed Roman survey of the Galactic Plane, with its NIR passbands and exquisite spacial resolution, promises groundbreaking insights for a wide range of time-domain galactic astrophysics.
In this white paper, we describe the scientific returns possible from the combination of the Roman Galactic Plane Survey with the data from LSST.

\section{Coordinating Roman and Rubin Surveys of the Galactic Plane}
\label{sec:intro}
During the next decade, ground-breaking multi-wavelength survey capabilities will be operating simultaneously: NASA's Nancy Grace Roman Space Telescope \cite{Akeson2019}, with its high-resolution NIR imager and the NSF's Vera C. Rubin Observatory 10\,yr optical survey of the southern sky \cite{Ivezic2019}.  
Contemporaneous observations of key regions of the Galactic Plane from both facilities can track the temporal evolution of stellar Spectral Energy Distributions (SEDs), dramatically enhancing our understanding of a wide range of stellar variability.  The complementary bandpasses also enable us to optimize the spatial regions surveyed by each facility, as Roman will be able to observe to greater depth in highly extincted regions, while Rubin's wider field of view enables it to survey a larger footprint.  By combining both datasets, we will give unprecedented insight into Galactic Structures.  Roman's high spatial resolution and precise astrometry will deliver astrometry in crowded regions that were inaccessible to ESA's {\em Gaia Mission} \cite{2015Ivezic}.    Roman and Rubin are also highly complementary in {\em time}.  Roman's Wide Field Imager can deliver high spatial resolution imaging in NIR passbands for a large region across the Milky Way Plane, while Rubin will provide long-baseline time-series at optical wavelengths.  By coordinating the times and intervals between observations between the surveys, we can maximize both the overall cadence and time series colour measurements.  
In Section~\ref{sec:science_goals} we discuss the wide range of science that can benefit from coordinating the observing strategies of these landmark surveys, reflecting contributions from across the Rubin Science Collaborations, particularly those for Transients and Variable Stars\footnote{https://lsst-tvssc.github.io/ \cite{Hambleton2023}} and Stars, Milky Way and Local Volume\footnote{https://rubin-smwlv.github.io/}. While combining many different science goals into a common survey strategy can be challenging, we build on prior work by our team within the Rubin Science Collaborations towards a common survey strategy for the Rubin Observatory's Galactic Plane survey.  In Section~\ref{sec:strategy} we discuss potential strategies for Roman observations of this region and Section~\ref{sec:metrics} outline potential metrics that can be used in the evaluation of alternative strategies. 

\section{Scientific Motivations}
\label{sec:science_goals}
\noindent{\bf Black Hole (BH) and Planetary Microlensing:} 
Gravitational microlensing occurs when a foreground object (lens) passes directly between the observer and a luminous background source. The gravity of the lens deflects light from the source causing the source to brighten and fade as it moves into and out of alignment with the lens, with a timescale, $t_{E}$ that is proportional to the lens mass (months to years for BHs, fraction of hours to days for planets). These events will be observed by LSST; however, to fully characterize them and reveal the true nature of the lens, an independent measurement of spectral type and source distance is needed.
A single visit by Roman to the Galactic Plane fields observed by the LSST will allow lens flux analysis and constrain possible scenarios. This idea is outlined and explored in Science Pitch by Bachelet et al. If possible, return visits to regions of special scientific interest after several years would better constrain the model parameters (see Section~\ref{sec:strategy}).

\noindent{\bf Microlensing optical depth:} 
Optical depth for microlensing provides a probability that a given star will be lensed, and is critical to understanding the true population of lensing objects \cite{1994Kiraga}.
While the optical depth has been measured for the Bulge \cite{2003Afonso, 2013Sumi, 2019Mroz}, the detections are too sparse across the rest of the Plane \cite{2009Rahal, 2017Moniez, 2020Mroz}.  
LSST will overcome this deficit, but its data has to be corrected for blending and extinction. Observing the Galactic Plane with high-angular resolution in the IR with Roman will produce a reference catalogue of stars, which combined with the LSST time-series data to more accurately correct these effects.


\noindent{\bf Predicted Lensing:} Time-domain observations in the Galactic Plane with Roman would produce better constraints on parallax and proper motion of stars in obscured or crowded regions, for which Gaia was not designed. This could lead to better prediction of future microlensing events \cite{2014Sahu, NeilsenBramich2018, 2018McGill, 2022Kluter}. In the past, these results have been used for mass measurements of Proxima Centauri (\cite{2018Zurlo}) or for testing the mass-radius relation of white dwarfs (\cite{2023McGill}).


\noindent{\bf Non-interacting Compact Object Binaries:} 
An exciting new population of wide-period, non-interacting compact object binaries has begun to be uncovered recently through serendipitous spectroscopic and wide-area astrometric surveys \citep{thompson19, gaiabh1, gaiabh2}. These have recently revealed the most massive stellar-mass black hole in the Milky Way, $M_{\rm BH}$\,$\approx$\,33\,M$_{\odot}$ \citep{gaiabh3}. Unless they originate in the halo, these are expected to be located close to the Galactic plane. Given how quickly these (astrometric binaries) were discovered following the availability of non-single-star astrometric solutions suggests that they are likely the tip of the iceberg of a much larger population. Roman will have the astrometric sensitivity to find these \citep{gandhi24_roman}, while Rubin will allow detailed characterisation of the donor star and line-of-sight reddening. In cases of fortuitous inclination angles, eclipses and self-lensing should also be detectable and help to characterise the systems in detail \citep{selflensing}. 

\noindent{\bf Cataclysmic Variables (CVs):} 
CVs, close binaries with a white dwarf (WD) accreting from a late main-sequence star or brown dwarf, represent the most common end-product of binary star evolution, and their frequency of occurrence places a constraint on stellar evolution models. However, an accurate estimate of the population has yet to be completed in the Galactic Plane due to their faintness and crowding. High cadence data allows their orbital periods to be measured, testing the predictions of a minimum period $P\sim $80\,min \citep{GoliaschNelson2015}, while longer-baseline data will detect dwarf novae in outburst.  Their blue colours mean that LSST data will complement the red/NIR data from Roman, while both surveys will push $\sim$3\,mag fainter than the OGLE survey \citep{Mroz2015}. Moreover, ZTF data revealed a larger number of CVs than previous out-of-plane surveys \citep{Szkody2021}. The simultaneous LSST and Roman data will be especially important for finding the many currently missing CVs with tightly magnetic WDs.

\noindent{\bf RR~Lyrae:}
Old, metal-poor stars in the inner Galaxy play a crucial role in understanding the formation of the Galactic bar and Bulge. Among these ancient stars, RR Lyrae stars (RRL) stand out as important both as population tracers and distance indicators. LSST will reach the far side of the Bulge, discovering new variables and determining their periods with extreme precision, extending well beyond the Galactic Center \cite{2024diCriscienzo}. However, the crowding limit hinders the reconstruction of light curves necessary to determine the average magnitude useful for deriving distance, reddening, and metallicity. The Roman telescope would offer the possibility to complement the work done by Rubin deriving accurate NIR mean magnitudes.

\noindent{\bf Galactic Structure: } 
Synergies between Roman and Rubin within the Bulge region will provide critical insights into the Milky Way structure, kinematics and evolution \citep{street24_roman}.  The Galactic bar extends out into the plane between longitudes, $-11^\circ \lesssim l \lesssim 30^\circ$ \citep{wegg15} (beyond the scope of the Bulge surveys), where Roman and Rubin plane surveys can characterize stellar populations, including variables.  The use of NIR period-metallicity-luminosity relations \citep{marconi15} will allow the precise determination of distances, while extinction maps will be crucial for determining the structure of the Milky Way from the distribution of stellar populations \citep{prudil24}.  Particularly using Classical Cepheids we can trace Milky Way warp and spiral arms.  Roman footprints overlapping previous IR surveys such as VVV/VVVx will facilitate new science with archive data and probe deeper into these regions.

\noindent{\bf Star Forming Regions} 
Regions with ongoing star formation, also called star-forming regions (SFRs), are associated with Nebulae, H\textsc{ii} region, Dark clouds, dust and
Giant Molecular Clouds. They are mainly found in the spiral arms of the Galactic Plane. 
A combined observational strategy leveraging deep photometry and astrometry, both in the optical and NIR bands with Rubin LSST \citep{prisi23} and Roman observations would revolutionize our understanding of 1) the first phases of stellar evolution; 2) cluster evolution and fate; 3) global properties of the Galactic structures, including the substructures such as the Radcliffe Wave \citep{alves20}, and others \cite{kuhn23}.
A combined observational strategy including wide field capabilities can be crucial to ascertain the validity of structures such as galactic strings \cite{kounkel19,jerabkova19,beccari20,wang22}, thereby offering new insights into the fundamentals of star formation.

 \noindent{\bf Young Stellar Objects: } 
Young stellar objects (YSOs) are characterized by photometric variability at all the timescales, from short- (hours), mid- (days, months), and long-term (years) timescales, their variability being part of the definition of classical T Tauri stars (CTTSs; \citep{joy45}).
YSOs are complex systems that consist of several components: a central forming star, surrounding material or disk from which material is accreted onto the stellar surface, bipolar jets ejected supersonically into the surrounding medium and causing the formation of shocks. All of these components emits a wide range of wavelengths, and IR observations are crucial to detect more embedded sources and knots in stellar jets.  This complements data from optical/UV emission which is fundamental to investigate the mechanisms at work in accreting young stars.
Simultaneous Rubin LSST and Roman data will be crucial for the characterization of such complex systems.

\noindent{\bf Neptune Trojans: } 
Trojans are asteroids that orbit at the stable Lagrangian points, 60$^{\circ}$ ahead and behind the major planets; currently, 22 are known orbiting with Neptune \cite{MPC}.  Uncovering their true population would help us to understand their origins and dynamics, but they are brighter in the NIR thanks to reflection and emission than the optical ($r$\,$\sim$22.5-24.5) with relatively low proper motions ($\sim$2$^{\circ}$ year$^{-1}$). During both Roman and Rubin's phases of main science operations, one of the Trojan groups will lie close to the Galactic Plane, of whom $\sim$10 are expected to be brighter than $r$=22.5mag. A single visit by Roman to this region to measure the NIR reflectance spectra of these objects significantly improves their characterisation.  A return visit to the same region after a short interval ($\sim$day) would also enable their astrometric motion to be measured. 

\noindent{\bf Bulge Globular clusters (GCs): } Studying the old stellar populations in GCs provides insight into the dynamical and physical conditions present during the Milky Way's early phases of star formation \cite{2003Krauss, 2014Roediger}. Measuring the periods, colours and luminosities of RR~Lyrae in these clusters will provide accurate distances.  Unfortunately, many of the galactic GCs lie in fields with high reddening and differential extinction, which has limited the precision possible from previous surveys.  Deep NIR imaging from Roman can overcome these issues, in combination with contemporaneous optical time-series from Rubin. These results can then be compared with similar studies of halo clusters and used to calibrate models of Galactic formation and evolution \citep{2017Binney}.

\section{Survey Strategies}
\label{sec:strategy}
By coordinating Roman's survey strategy across the Galactic Plane with Rubin's observations of the same region, we can serve the needs of a wide range of science with a single visit per field. If both optical and NIR surveys imaged the region contemporaneously, they would deliver not only high-resolution data needed for de-blending, but would also derive the fill SED of all observed sources.
The Rubin survey footprint in this region (Fig.~\ref{fig:rubin_footprint}) was determined by the community recommendation \cite{Street2023} and highlights the number of regions of special scientific interest. Coordinated imaging by Roman across this region offers enormous scientific returns for the topics described in Section~\ref{sec:science_goals}.
We propose to use F087 and F213 filters to constrain the SEDs best.
A single filed observation would require 985.28~seconds, assuming that the exposure time for each filter is 55~seconds, and one visit requires five dithers (slew time of 23.4~seconds). If the survey would require only a small angle shift between fields (slew time of 54.8~seconds), the total covered field would be 718.5 square degrees, which is less than half of the total area of Rubin's Galactic Plane coverage. Roman Galactic Plane survey could focus fields, such as Far Bulge, which is not included in the Galactic Bulge Survey, or selected fields of the Galactic Disc.

If time permits, we also highlight the value of obtaining second-epoch observations with Roman, also coordinated with the later phases of Rubin's LSST.  Roman's precision astrometry would provide a wealth of astrometric and proper motion measurements for regions where these data cannot be provided by Gaia.  This is especially valuable for microlensing and in clusters and SFRs.  





\section{Metrics}
\label{sec:metrics}
\begin{figure}
    \centering
    \includegraphics[width=16cm]{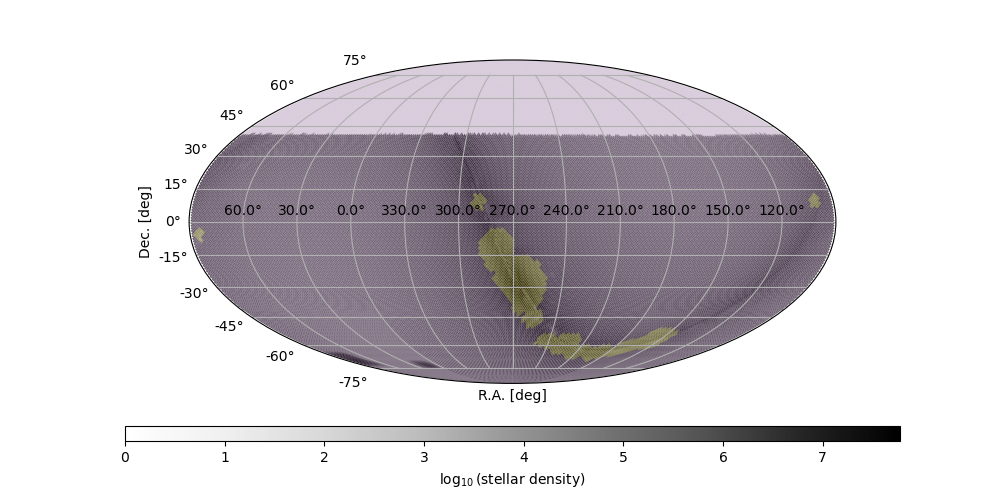}
    \caption{Outline of the Rubin Galactic Plane survey footprint (yellow overlay) plotted over a map of stellar density. This survey footprint is highly complementary to that proposed for the Roman Galactic Plane survey.}
    \label{fig:rubin_footprint}
\end{figure}

The primary metric to evaluate the scientific return of Roman observations for the science goals described above will measure the spatial overlap between those observations and the region included in Rubin's Galatic Plane survey footprint (Fig.~\ref{fig:rubin_footprint}).  Observations of this region at any time will be valuable, but they will be particularly valuable if they occur contemporaneously with those of Rubin.  For this reason, we also recommend metrics to measure the average time-gap between Roman and Rubin observations within this survey footprint.  Last but not least, these metrics should track the intervals between contemporaneous (ideally within 24hrs) observations in filters that are widely separated in wavelength which will provide the best measurement of stellar SED, particularly Roman's F213, F087 and Rubin's SDSS-$g, r$ or $i$.  

Over the last few years, Rubin Observatory and the worldwide science community have conducted an innovative drive to design and build metrics to describe the impact of survey strategy changes on different science goals.  Our team designed many metrics relating to galactic science (including survey region, temporal coverage and use of filters), and contributed Python code to the open-source Rubin Metric Analysis Framework \cite{Jones2014}.  This code is open source and can be re-used, or adapted into Roman's own survey simulation framework.  Work is in progress on metrics that evaluate how changes to the timing of Roman and Rubin observations in the Bulge affect the analysis of microlensing events (Verala et al. in prep.).  Code to evaluate the spatial overlap between the survey footprints of both missions has also been trialed, and is available for adaptation.  

The Rubin Galactic Plane footprint was arrived at by combining the survey regions of interest to a number of different science cases \cite{Street2023}.  Roman can accomplish a great deal of science with a single visit to all fields within the Galacic Plane, but if time permits a later revisit to at least some of this region, then a process similar to Rubin's could be used to identify regions of high scientific priority.  

\end{document}